\documentclass[reprint,
superscriptaddress,
 amsmath,amssymb,
 aps,
]{revtex4-2}

\usepackage{graphicx}
\usepackage{caption}
\usepackage{subcaption}
\usepackage{dcolumn}
\usepackage{bm}

\usepackage[utf8]{inputenc}
\usepackage[T1]{fontenc}
\usepackage{mathptmx}
\usepackage{etoolbox}
\usepackage{xcolor}
\usepackage{adjustbox}
\usepackage{float}

\begin{document}

\preprint{APS/123-QED}

\title{Optimizing Hybrid Ferromagnetic Metal-Ferrimagnetic Insulator Spin-Hall Nano-Oscillators: A Micromagnetic Study} 

\author{Robert Xi}
\affiliation{Center for Quantum Phenomena, Department of Physics, New York University, New York, New York 10003, USA}

\author{Ya-An Lai}
\affiliation{Center for Quantum Phenomena, Department of Physics, New York University, New York, New York 10003, USA}

\author{Andrew D. Kent}
\affiliation{Center for Quantum Phenomena, Department of Physics, New York University, New York, New York 10003, USA}

\date{\today}

\begin{abstract}
Spin-Hall nano-oscillators (SHNO) are nanoscale spintronic devices that generate high-frequency (GHz) microwave signals useful for various applications such as neuromorphic computing and creating Ising systems. Recent research demonstrated that hybrid SHNOs consisting of a ferromagnetic metal (permalloy) and lithium aluminum ferrite (LAFO), a ferrimagnetic insulator, thin films have advantages in having lower auto-oscillation threshold currents ($I_{\text{th}}$) and generating larger microwave output power, making this hybrid structure an attractive candidate for spintronic applications.  It is essential to understand how the tunable material properties of LAFO, {\emph e.g.}, its thickness, perpendicular magnetic anisotropy ($K_u$), and saturation magnetization ($M_s$), affect magnetic dynamics in hybrid SHNOs. We investigate the change in $I_{\text{th}}$ and the output power of the device as the LAFO parameters vary. We find the $I_{\text{th}}$ does not depend strongly on these parameters, but the output power has a highly nonlinear dependence on $M_s$ and $K_u$. We further investigate the nature of the excited spin-wave modes as a function of $K_u$ and determine a critical value of $K_u$ above which propagating spin-waves are excited. Our simulation results provide a roadmap for designing hybrid SHNOs to achieve targeted spin excitation characteristics.
\end{abstract}


\maketitle 

\section{Introduction}
Spin-Hall nano-oscillators (SHNO) have been of great research interest in achieving highly efficient spintronic devices for applications in the field of magnonics~\cite{Khitun2010,Han2019,Demidov2020} and other novel information processing architectures such as Ising systems~\cite{wang2019oim,makiuchi2021parametron,elyasi2022stochasticity}, data communication~\cite{Sharma2021,Litvinenko2022} and neuromorphic computing \cite{Locatelli2014,Torrejon2017,Tsunegi2019,Zahedinejad2019,Romera2022,Marković2022}. SHNO typically consists of a heavy metal (HM) layer and a ferromagnetic layer (FM) in which spin orbit torque is generated by the spin-Hall effect in the HM layer that drives magnetic moments to auto-oscillation in the FM layer~\cite{Kurebayashi2010,balinsky2016,safranski2017,Evelt2018}. Insulating ferrimagnetic layers have been shown to effectively transmit spin angular momentum due to their low magnetic damping, associated with their lack of conduction electrons~\cite{Kajiwara2010}.

Recent studies have developed a new class of magnetic insulating thin films, lithium aluminum ferrite (LAFO), which possesses desirable properties for SHNO integration such as tunable magnetic anisotropy, magnetization, low damping and absence of dead magnetic layers~\cite{Zheng2024}. A recent study demonstrated that a hybrid SHNO configuration consisting of a permalloy/platinum bilayer nanowire (Py/Pt) on top of an extended LAFO thin film enhances microwave power emission and leads to excitation of localized spin-wave edge modes~\cite{Ren2023}. With the tunability of LAFO properties, it is essential to understand how the thickness, saturation magnetization ($M_s$), and magnetic anisotropy ($K_u$) of LAFO affect the auto-oscillation threshold currents and the amplitude of the modes. More importantly, as propagating spin waves can be excited in SHNO consisting of materials with perpendicular magnetic anisotropy (PMA)~\cite{evelt2018emission,chen2023self}, it is also essential to investigate the role of PMA in hybrid SHNO.

In this study, we use micromagnetic methods to study the effect of LAFO thickness, $M_s$, and $K_u$ on SHNO characteristics. We find the critical current for the onset of auto-oscillations to be relatively insensitive to magnetization and perpendicular magnetic anisotropy. However, as the perpendicular magnetic anisotropy increases, the nature of the excited spin waves changes from localized bulk and edge modes to propagating spin wave modes and then to localized droplet modes. Our simulation findings thus provide insights into the design of hybrid SHNOs to obtain desired spin excitation characteristics.
\begin{figure*}
    \centering
    \includegraphics[width=0.9\linewidth]{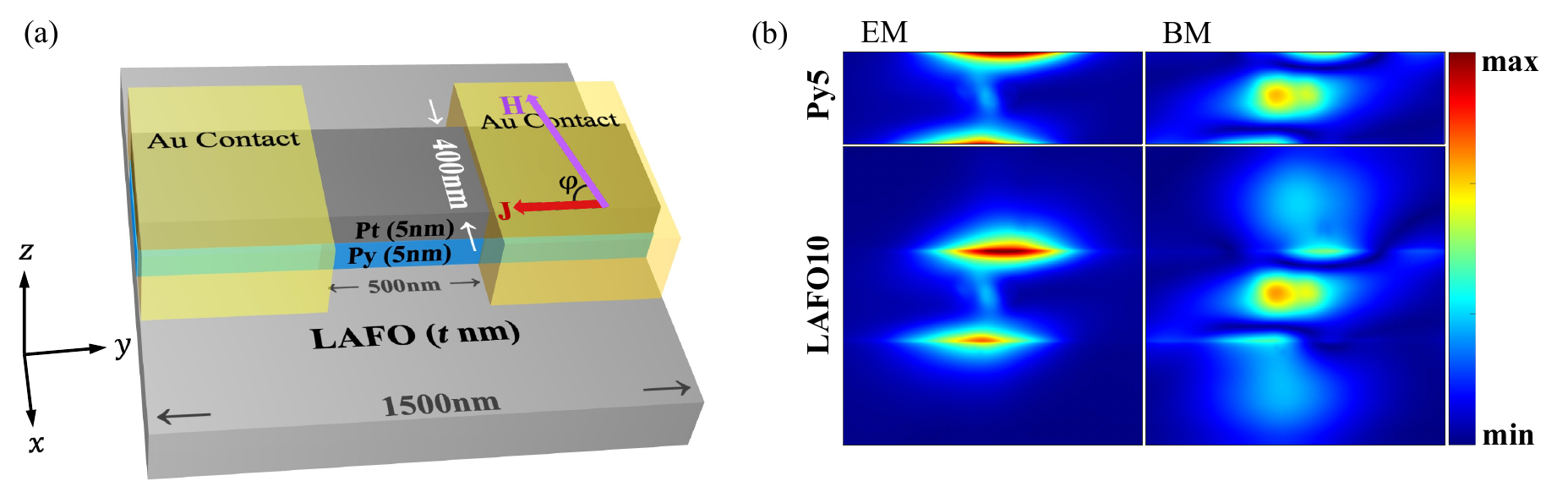}
    \captionsetup{labelfont=bf}
    \caption{\textbf{Hybrid nano-oscillator and modes spatial FFT images. (a)} Schematic of a hybrid SHNO. The device consists of LAFO ($t$)/Py5/Pt5. The external field is applied at an angle $\phi$ to the current, and the applied charge current is restricted to the center region of the Py nanowire, indicated in a dark blue color. \textbf{(b)} Spatial FFT profiles of the edge mode (EM) and bulk mode (BM) in the Py nanowire and LAFO (10 nm) layer. An in-plane external field $\mu_0H_{\mathrm{ext}}=0.08$ T is applied at $\phi=70^\circ$. } 
    \label{fig:schematic}
\end{figure*}

\begin{figure*}
    \centering
    \includegraphics[width=0.9\linewidth]{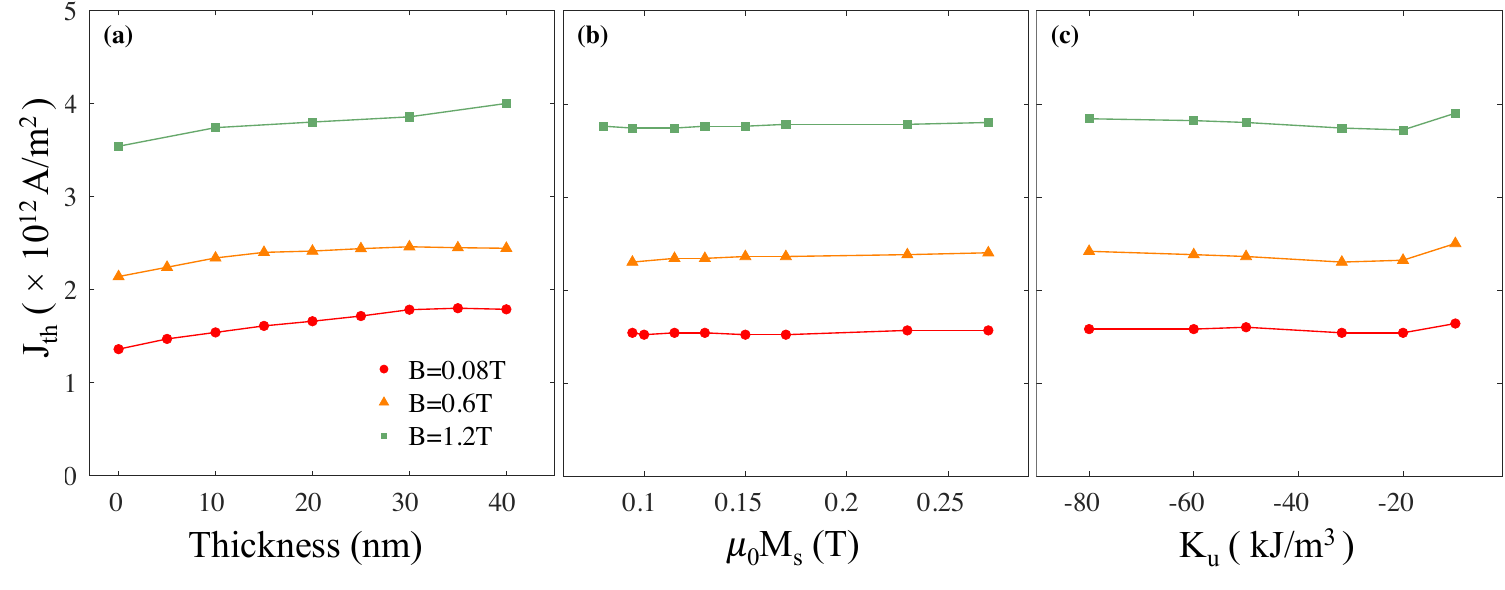}
    \captionsetup{labelfont=bf}
  \caption{\textbf{Threshold Current Density $J_\mathrm{th}$ as Function of }\textbf{(a)} LAFO thickness ($\mu_0 M_{s\mathrm{, LAFO}}=0.0942$ T, $K_u=-31.7$ kJ/m$^3$), \textbf{(b)} LAFO saturation magnetization $M_{s\mathrm{,LAFO}}$ ($t_\mathrm{LAFO} = 10$ nm, $K_u=-31.7$ kJ/m$^3$), and \textbf{(c)} LAFO perpendicular anisotropy constant $K_\mathrm{u}$ ($t_\mathrm{LAFO} = 10$ nm, $\mu_0M_{s\mathrm{,LAFO}}=0.0942$ T) with an in-plane external field  $\mu_0H_{\mathrm{ext}}=0.08,\;0.6$ and $1.2$ T at $\phi=70^{\circ}$. For all the simulations, $\alpha_\mathrm{LAFO}=1\times10^{-3}$, $\alpha_\mathrm{Py}=2.6\times10^{-2}$,  $\mu_0M_\mathrm{eff,Py}=0.78$ T, and  $\mu_0M_{s\mathrm{,Py}}=1.08$ T.}
    \label{fig:Jth}
\end{figure*}

\begin{figure*}[htbp]
  \centering
  \includegraphics[width=0.9\linewidth]{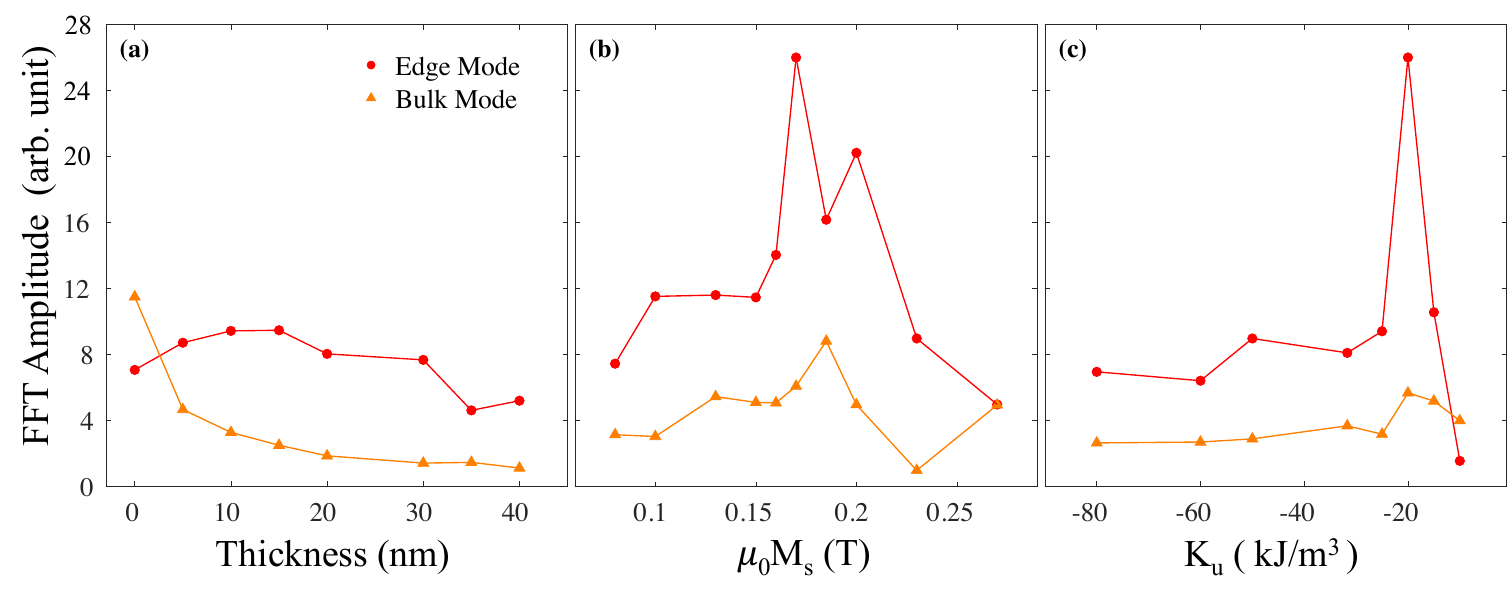}
  \captionsetup{labelfont=bf}
  \caption{\textbf{Edge Mode and Bulk Mode Output Power as function of } \textbf{(a)} LAFO thickness. \textbf{(b)} LAFO saturation magnetization $\mu_0 M_{s\mathrm{,LAFO}}$ ( $t_\mathrm{LAFO} = 20$ nm, $K_u=-31.7$ kJ/m$^3$). \textbf{(c)} LAFO perpendicular anisotropy constant $K_u$ ($t_\mathrm{LAFO} = 10$ nm, $\mu_0M_{s\mathrm{,LAFO}}=0.0942$ T). All simulations were conducted with an in-plane $\mu_0H_{\mathrm{ext}}=0.08$ T at $\phi=70^{\circ}$ with constant parameters $\alpha_\mathrm{LAFO}=1\times10^{-3}$, $\alpha_\mathrm{Py}=2.6\times10^{-2}$, $\mu_0 M_\mathrm{eff, Py}=0.78$ T, and $\mu_0M_{s\mathrm{,Py}}=1.08$ T. The current density is set to $J_e=1.2J_\mathrm{th}$.}
  \label{fig:output}
\end{figure*}

\section{Micromagnetic Simulations}
We used the MuMax3 micromagnetic solver~\cite{Vansteenkiste2014} to simulate an LAFO/Py/Pt hybrid SHNO device. In the geometry setup, a $400\times1500\times5$ nm$^3$ Py nanowire is on top of a $1500\times1500\times t$ nm$^3$ ($t=0-40$ nm) LAFO layer, as shown in Fig.~\ref{fig:schematic}(a). The simulation employs a cell size of $5\times 5\times 5$ nm$^3$, which is smaller than the exchange length of Py and LAFO. Py parameters listed below are kept constant throughout simulations, $\mu_0 M_{s\mathrm{, Py}}=1.08$ T obtained from commonly used value from references, an effective magnetization $\mu_0 M_{\mathrm{eff, Py}} = 0.78$ T, and a Gilbert damping constant $\alpha_{\mathrm{Py}}=2.6\times 10^{-2}$ from ferromagnetic resonance (FMR) measurements~\cite{Ren2023}. The exchange constant between the Py layer and LAFO layer is taken to be half of the harmonic mean of the two layers. To reduce the spin-wave reflection at the boundary, we set an exponentially increased damping approaching the boundaries. Periodic boundary conditions are implemented in the $x$ direction to properly model the demagnetization field.

To simulate auto-oscillation excitation, we apply a charge current density $J_e$ to the central region of the Py nanowire because this is where $J_e$ flows through in the device as the remaining Py region is covered by electrical contacts. The Oersted field generated from $J_e$ is negligible and not included in our study. The spin current density $J_s$ is calculated from $J_e$ using the relation $J_s=\theta_\mathrm{SH}J_e$, where $\theta_\mathrm{SH}$ is the spin Hall angle of Pt and is set to 0.15. $J_e$ is slowly increased at a rate of $4\times 10^8 \mathrm{A/m}^2\mathrm{s}$ so that the system reaches a steady state at each time step. We determine the threshold current density $J_\mathrm{th}$, which is the minimum $J_e$ to excite the auto-oscillation state, by monitoring whether the average maximum torque acting on the Py magnetization increases within 150 ns simulation time. An increase in the torque implies that the Py magnetization is driven out of equilibrium and into auto-oscillation. In our simulation studies, we set $J_e=1.2J_\mathrm{th}$ to eliminate $J_\mathrm{th}$-dependent effects. To resolve the resonant frequencies of the auto-oscillation state, we performed the fast Fourier transform (FFT) on the time evolution of the spatially averaged steady-state magnetization along the $z$ direction $\bar{m}_z(t)$ in the central Py nanowire. We find the auto-oscillation frequencies by fitting the resulting FFT amplitude spectrum to a Lorentzian function. Pixel-wise spatial FFT is conducted to obtain the spatial profile for each oscillation mode in the Py nanowire and LAFO layer following the FFT method of Ref.~\cite{McMichael2005}. 
To simulate FMR experiments, an in-plane perturbing magnetic field in the form of a sinc function of magnitude 0.5mT is applied to sample with a fixed external field ($H_\mathrm{ext}=0.08T$). Then, we perform FFT on the excited $\bar{m}_z(t)$ to determine the FMR frequencies.

\section{Results and discussion}

Previous work has shown that the main auto-oscillation modes in LAFO/Py/Pt hybrid SHNO devices are an edge mode (EM) and a bulk mode (BM)~\cite{Ren2023}. For EMs, the auto-oscillation occurs at a lower frequency with a higher oscillation amplitude near the edges of the Py nanowire, where the internal magnetic field is reduced due to the demagnetizing field. For the BM, the auto-oscillation occurs at high frequency with a high oscillation amplitude near the center of the Py nanowire, as shown in Fig.~\ref{fig:schematic}(b). 

\subsection{Threshold Current}
Lowering the threshold current to drive auto-oscillation is crucial to minimize power consumption and heat generation, which affect the thermal stability of SHNO devices. Therefore, we first examine the effect on $J_\mathrm{th}$ as LAFO thickness ($t_\mathrm{LAFO}$), magnetization ($M_{s\mathrm{,LAFO}}$), and perpendicular magnetic anisotropy ($K_{u}$) vary at $\mu_0H_{\mathrm{ext}}=0.08, \;0.6$, and $1.2$T. A macrospin model~\cite{Ren2023} predicts that $J_\mathrm{th}$ is proportional to $t_\mathrm{LAFO}$, $M_{s\mathrm{,LAFO}}$, and $K_{u}$: 
\begin{equation}
    J_\mathrm{th} \propto \left(\alpha_\mathrm{Py} M_{s\mathrm{, Py}} t_\mathrm{Py}+\alpha_\mathrm{LAFO} M_{s\mathrm{,LAFO}} t_\mathrm{LAFO}\right) \bar{M}_\mathrm{eff},
    \label{Eq:Jth}
\end{equation}
where 
\begin{equation}
\bar{M}_{\mathrm{eff}}=\frac{t_{\mathrm{Py}}M_{\mathrm{eff,Py}}M_{s\mathrm{,Py}}+t_{\mathrm{LAFO}}M_{\mathrm{eff,LAFO}}M_{s\mathrm{,LAFO}}}{t_{\mathrm{Py}}M_{s\mathrm{,Py}}+t_{\mathrm{LAFO}}M_{s\mathrm{,LAFO}}}
    \label{Eq:aveMeff}
\end{equation}
is the average effective magnetization of the hybrid nano-oscillator, and the effective magnetization for each layer is calculated by $M_\mathrm{eff}=M_s-2K_{u}/\mu_0M_s$.
Figure~\ref{fig:Jth}(a) shows that $J_\mathrm{th}$ slowly increases with $t_\mathrm{LAFO}$, which agrees with the macrospin model prediction. However, $J_\mathrm{th}$ does not depend noticably on $M_{s\mathrm{,LAFO}}$ and $K_{u}$, as seen in Fig.~\ref{fig:Jth}(b) and (c). We attribute this to the fact that $M_{s\mathrm{,LAFO}}$ is only one-tenth that of $M_{s\mathrm{,Py}}$, leading to a negligible effect on $J_\mathrm{th}$, as expected from Eq.~\ref{Eq:Jth}. From our micromagnetic study, we conclude that in the regime of $0.08$T$<\mu_0M_{s\mathrm{,LAFO}}<0.25$T and $0.35$T$<\mu_0M_{\mathrm{eff,LAFO}}<2.59$ T, reducing $t_{\mathrm{LAFO}}$ is the most effective means to lower $J_\mathrm{th}$.

\begin{figure}[htbp]
    \centering
    \includegraphics[width=0.7\linewidth]{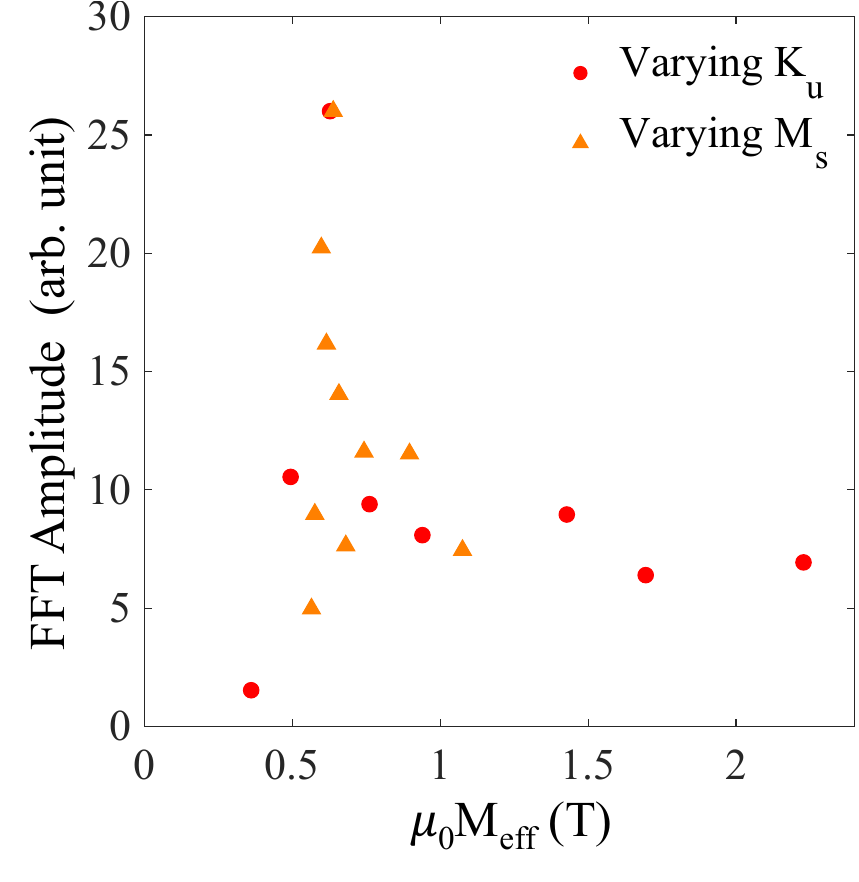}
    \captionsetup{labelfont=bf}
    \caption{\textbf{Edge Mode Output Power as Function of LAFO Effective Magnetization $M_\mathrm{eff,LAFO}$}.  The effective magnetization is calculated for two cases. First, fix $\mu_0M_s=0.0942$ T and vary $K_u$ (red circles). Second, fix $K_u=-31.7\text{ kJ/m}^3$ and vary $M_{s,\mathrm{LAFO}}$ (orange triangles). The external field is fixed at $\mu_0H_{\mathrm{ext}}=0.08$ T at $\phi=70^{\circ}$.} 
    \label{fig:EM amp vs Meff}
\end{figure}

\subsection{Output Power}
 It is also crucial to maximize the output power, as it allows better detection of electrical signals in applications. Research has demonstrated that the incorporation of an extended LAFO thin film beneath a Pt/Py nanowire can significantly enhance power emission~\cite{Ren2023}. However, the relationship between this enhanced power emission and LAFO parameters is yet to be explored. Now, we consider how the output power varies with the LAFO parameters ($t_\mathrm{LAFO}$, $M_{s\mathrm{, LAFO}}$, and $K_{u}$). The auto-oscillation amplitude is shown as a function of these parameters in Fig.~\ref{fig:output}. Figure~\ref{fig:output}(a) shows that the output power of the EM does not depend strongly on $t_\mathrm{LAFO}$, whereas the output power of the BM decreases as $t_\mathrm{LAFO}$ increases. 

\begin{figure*}
  \centering
  \includegraphics[width=\linewidth]{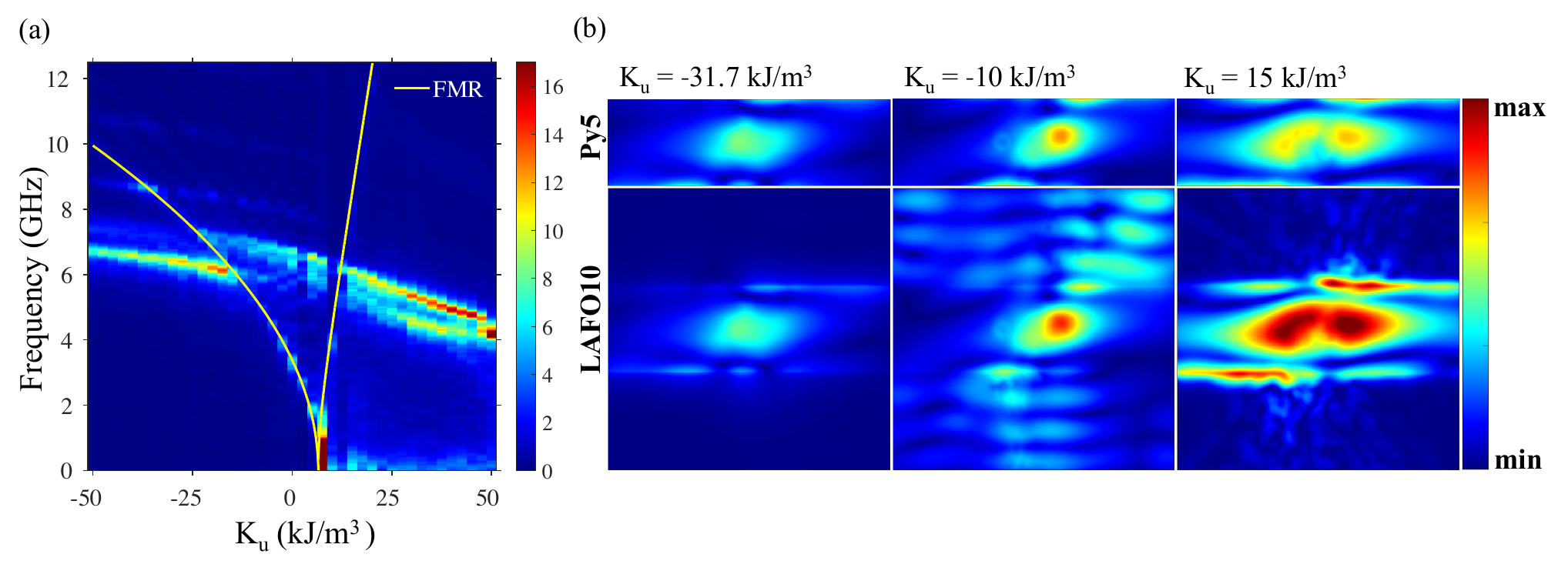}
  \captionsetup{labelfont=bf} 
  \caption{\textbf{PSD and Spatial Profiles from Micromagnetic Modeling.} \textbf{(a)} PSD map as a function of frequency and $K_{u}$ with $\mu_0H_{\mathrm{ext}} = 0.08$ T at $\phi = 70^{\circ}$, and $J=2\times10^{12}$ A/m$^2$, $\mu_0M_{\mathrm{s}} = 0.0942$T for LAFO10/Py5. The yellow curve indicates the FMR frequency. \textbf{(b)} Spatial FFT images of the LAFO layer and Py layer at the BM auto-oscillation frequency for $K_{u} = -31.7,\, -10,$ and $ 15$ kJ/m$^3$. The linear color scale is on the right, where the color represents the FFT amplitude of the magnetization along the $z$ direction.}
  \label{fig:PSD spin excite+propagation}
\end{figure*}

\begin{figure}[htbp]
  \centering
  \includegraphics[width=0.8\linewidth]{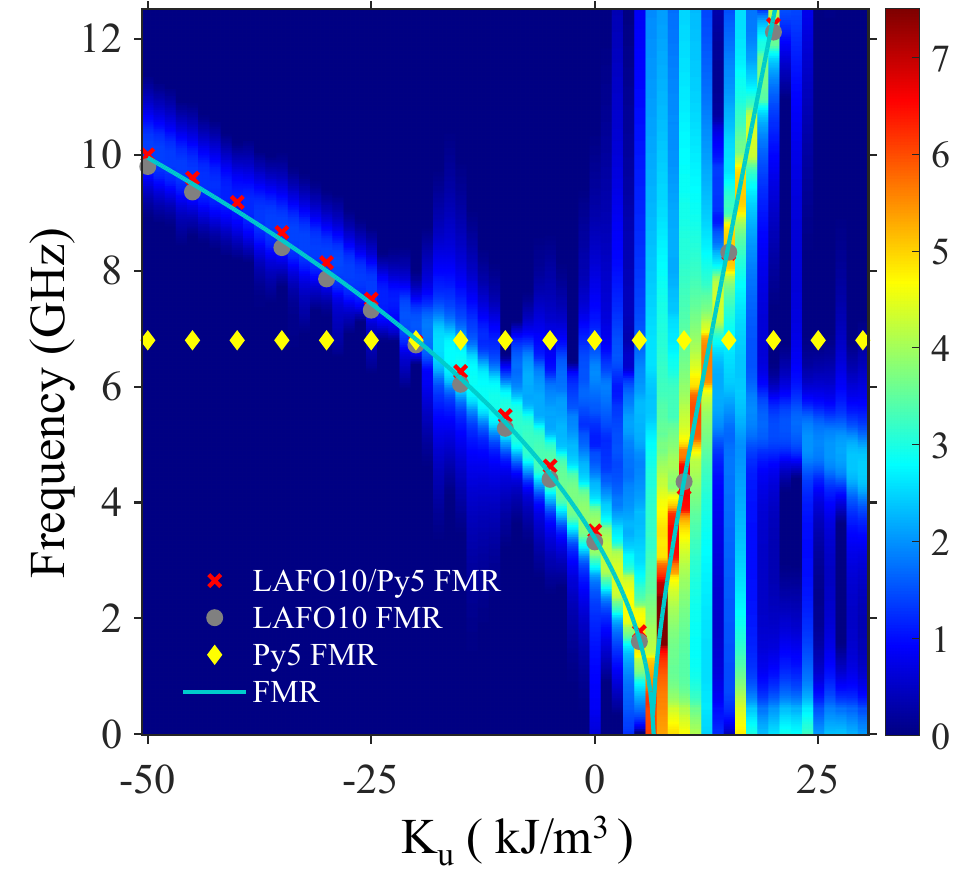} 
  \captionsetup{labelfont=bf} 
  \caption{\textbf{FMR Simulation.} PSD of LAFO10/Py5/Pt5 as a function of $K_u$ at a fixed field $\mu_0H_{\mathrm{ext}} = 0.08$ T at $\phi = 70^{\circ}$ and fixed saturation magnetization $\mu_0M_{s\mathrm{,LAFO}}=0.0942$ T and $\mu_0M_{s\mathrm{,Py}}=1.08$ T. The Kittel model with $\gamma=180$ GHz/T is shown by the blue curve. The FMR frequencies of LAFO10/Py5/Pt5, bare LAFO10, and bare Py5 are shown by red crosses, gray circles, and yellow squares, respectively.}
  \label{fig:FMR PSD}
\end{figure}
 
 Interestingly, Fig.~\ref{fig:output}(b) shows that there is a peak in the output power of EM and BM around $\mu_0 M_{s\mathrm{,LAFO}}=0.17$ T. 
 In Fig.~\ref{fig:output}(c), a peak in EM output power is observed around $K_u=-20$kJ/m$^3$. As $K_u$ becomes less negative (\emph{i.e.}, in-plane magnetization is less favored) and $M_{\mathrm{eff,LAFO}}$ decreases. This is advantageous for aligning the moments at the edge of the Py nanowire, generating a larger precession angle. To better understand why there is a maximum in the output power of the EM, $\mu_0M_\mathrm{eff,LAFO}$ is plotted against the EM output power in Fig.~\ref{fig:EM amp vs Meff} based on data from Fig.~\ref{fig:output}(b) and (c). We observe a peak in the EM output power around $\mu_0 M_\mathrm{eff,LAFO}=0.6$ T when varying either $K_u$ or $M_{s,\mathrm{LAFO}}$. This suggests that the peak is determined by the effective magnetization $M_\mathrm{eff,LAFO}$, rather than by $K_u$ or $M_{s\mathrm{,LAFO}}$ independently.

\subsection{Excited Mode Profiles} 
We now investigate the nature of the auto-oscillation modes as $K_{u}$ is varied. Figure~\ref{fig:PSD spin excite+propagation}(a) shows the PSD map as a function of frequency and $K_{u}$. The yellow curve is the LAFO FMR frequency calculated using the Kittel relation $f=\mu_0\gamma/(2\pi)[(H_\mathrm{ext}\sin\theta_M-M_{\mathrm{eff}}\cos2\theta_M)(H_\mathrm{ext}\sin\theta_M-M_{\mathrm{eff}}\cos^2\theta_M)]^{1/2}$, where $\gamma = 180$ GHz/T is the gyromagnetic ratio, $\mu_0$ is the vacuum permeability and $\theta_M$ is the equilibrium angle of the LAFO magnetization and is calculated by solving the angle that minimizes the energy. The PSD map is sectioned into three regions by the FMR curve, where the middle region has spin wave modes excited at frequencies higher than the FMR frequency $f_{\mathrm{FMR}}$ and the other regions have spin wave modes excited below $f_{\mathrm{FMR}}$. This indicates that the spin-wave modes are localized for both large negative and positive $K_u$ but can propagate for intermediate values of the perpendicular magnetic anisotropy~\cite{succar2023}. This is seen directly by the spatial FFT analysis shown in Fig.~\ref{fig:PSD spin excite+propagation}(b). For large negative $K_u$, the spin waves are localized near the Py edges, where the demagnetization field creates a potential well for the spin waves, associated with a local minima in $M_\mathrm{eff}$. The applied field in the plane and the in-plane anisotropy increase the depth of the spin-wave potential well and lead to further mode localization~\cite{dvornik2018origin}. As $K_{u}$ approaches zero from negative values, magnetization has less preference to align along a specific orientation. This leads to a shallower potential well for spin waves with a smaller $M_\mathrm{eff}$. As $M_\mathrm{eff}$ decreases, the auto-oscillation frequency and $f_{\mathrm{FMR}}$ decrease until the auto-oscillation frequency is greater than $f_{\mathrm{FMR}}$, leading to propagating spin wave modes. For large positive $K_u$, the LAFO is perpendicularly magnetized. Therefore, increasing $K_u$ leads to a decreasing $M_\mathrm{eff}$ but an increasing $f_{\mathrm{FMR}}$.

To better understand the interplay between mode behavior and intrinsic properties of the device, a FMR spectroscopy simulation is conducted with varying $K_u$, with values ranging from $-50$ to $30$ kJ/m$^3$. Figure~\ref{fig:FMR PSD} compares the PSD FMR map of an LAFO10/Py5/Py5 SHNO device with the FMR condition derived from the Kittel relation (yellow curve), the FMR spectrum of LAFO10 (cyan crosses) and Py5 (red crosses). The single FMR mode indicates that the LAFO and Py layers are strongly coupled. The similarity in $f_{\mathrm{FMR}}$ between the LAFO layer and the device suggests a minimal impact of the Py layer on the hybrid SHNO FMR frequency. In comparison to the data presented in Fig.~\ref{fig:PSD spin excite+propagation}(a), it is observed that the EM's peak output power corresponds to the point where $f_\mathrm{FMR,Py}$ and $f_\mathrm{FMR,LAFO}$ coincide. This observation suggests that the observed EM maximum output power is attributable to a resonance effect between the two layers. It should be noted that in Fig.~\ref{fig:FMR PSD} the amplitude of the device's FMR response shows a stepwise increase when $f_\mathrm{FMR,Py}$ exceeds $f_\mathrm{FMR,LAFO}$ around $K_u=-20 $ kJ/m$^3$. This is likely due to the fact that when $f_\mathrm{FMR,LAFO}>f_\mathrm{FMR,Py}$, the Py magnon frequencies fall below the lowest frequency LAFO magnons, impeding the propagation of magnons from the Py layer to the LAFO layer~\cite{fan2021}. In contrast, when $f_\mathrm{FMR,LAFO}< f_\mathrm{FMR,Py}$, the Py magnons can propagate into the LAFO layer. In addition, since the LAFO layer has an ultralow Gilbert damping, the FMR amplitude is enhanced.

\subsection{Summary}
In this work, we investigated the impact of various LAFO parameters on $J_\mathrm{th}$ and output power of the LAFO/Py/Pt hybrid SHNO using micromagnetic simulations. The results indicate that $J_\mathrm{th}$ is not significantly affected by changes in $t_\mathrm{LAFO}$, $M_{s,\mathrm{LAFO}}$, or $K_u$. However, a peak in the EM output power was observed when $M_{s,\mathrm{LAFO}}$ and $K_u$ are varied. Our findings reveal that the maximum EM output power occurs when $f_{\mathrm{FMR,LAFO}}$ and $f_{\mathrm{FMR,Py}}$ coincide. Our study also shows that the nature of BM changes between a localized mode and a propagating mode as $K_u$ varies. Our work provides a roadmap for the engineering of LAFO properties for building more efficient and powerful hybrid SHNO devices for future applications.

\begin{acknowledgments}
The authors acknowledge very helpful discussions with Haowen Ren and Yuri Suzuki. This research was supported by NSF DMR-2105114.
\end{acknowledgments}

\bibliography{biblio}

\begin{thebibliography}{28}%
\makeatletter
\providecommand \@ifxundefined [1]{%
 \@ifx{#1\undefined}
}%
\providecommand \@ifnum [1]{%
 \ifnum #1\expandafter \@firstoftwo
 \else \expandafter \@secondoftwo
 \fi
}%
\providecommand \@ifx [1]{%
 \ifx #1\expandafter \@firstoftwo
 \else \expandafter \@secondoftwo
 \fi
}%
\providecommand \natexlab [1]{#1}%
\providecommand \enquote  [1]{``#1''}%
\providecommand \bibnamefont  [1]{#1}%
\providecommand \bibfnamefont [1]{#1}%
\providecommand \citenamefont [1]{#1}%
\providecommand \href@noop [0]{\@secondoftwo}%
\providecommand \href [0]{\begingroup \@sanitize@url \@href}%
\providecommand \@href[1]{\@@startlink{#1}\@@href}%
\providecommand \@@href[1]{\endgroup#1\@@endlink}%
\providecommand \@sanitize@url [0]{\catcode `\\12\catcode `\$12\catcode `\&12\catcode `\#12\catcode `\^12\catcode `\_12\catcode `\%12\relax}%
\providecommand \@@startlink[1]{}%
\providecommand \@@endlink[0]{}%
\providecommand \url  [0]{\begingroup\@sanitize@url \@url }%
\providecommand \@url [1]{\endgroup\@href {#1}{\urlprefix }}%
\providecommand \urlprefix  [0]{URL }%
\providecommand \Eprint [0]{\href }%
\providecommand \doibase [0]{https://doi.org/}%
\providecommand \selectlanguage [0]{\@gobble}%
\providecommand \bibinfo  [0]{\@secondoftwo}%
\providecommand \bibfield  [0]{\@secondoftwo}%
\providecommand \translation [1]{[#1]}%
\providecommand \BibitemOpen [0]{}%
\providecommand \bibitemStop [0]{}%
\providecommand \bibitemNoStop [0]{.\EOS\space}%
\providecommand \EOS [0]{\spacefactor3000\relax}%
\providecommand \BibitemShut  [1]{\csname bibitem#1\endcsname}%
\let\auto@bib@innerbib\@empty
\bibitem [{\citenamefont {Khitun}\ \emph {et~al.}(2010)\citenamefont {Khitun}, \citenamefont {Bao},\ and\ \citenamefont {Wang}}]{Khitun2010}%
  \BibitemOpen
  \bibfield  {author} {\bibinfo {author} {\bibfnamefont {A.}~\bibnamefont {Khitun}}, \bibinfo {author} {\bibfnamefont {M.}~\bibnamefont {Bao}},\ and\ \bibinfo {author} {\bibfnamefont {K.~L.}\ \bibnamefont {Wang}},\ }\bibfield  {title} {\bibinfo {title} {Magnonic logic circuits},\ }\href@noop {} {\bibfield  {journal} {\bibinfo  {journal} {Journal of Physics D: Applied Physics}\ }\textbf {\bibinfo {volume} {43}},\ \bibinfo {pages} {264005} (\bibinfo {year} {2010})}\BibitemShut {NoStop}%
\bibitem [{\citenamefont {Han}\ \emph {et~al.}(2019)\citenamefont {Han}, \citenamefont {Zhang}, \citenamefont {Hou}, \citenamefont {Siddiqui},\ and\ \citenamefont {Liu}}]{Han2019}%
  \BibitemOpen
  \bibfield  {author} {\bibinfo {author} {\bibfnamefont {J.}~\bibnamefont {Han}}, \bibinfo {author} {\bibfnamefont {P.}~\bibnamefont {Zhang}}, \bibinfo {author} {\bibfnamefont {J.~T.}\ \bibnamefont {Hou}}, \bibinfo {author} {\bibfnamefont {S.~A.}\ \bibnamefont {Siddiqui}},\ and\ \bibinfo {author} {\bibfnamefont {L.}~\bibnamefont {Liu}},\ }\bibfield  {title} {\bibinfo {title} {Mutual control of coherent spin waves and magnetic domain walls in a magnonic device},\ }\href@noop {} {\bibfield  {journal} {\bibinfo  {journal} {Science}\ }\textbf {\bibinfo {volume} {366}},\ \bibinfo {pages} {1121} (\bibinfo {year} {2019})}\BibitemShut {NoStop}%
\bibitem [{\citenamefont {Demidov}\ \emph {et~al.}(2020)\citenamefont {Demidov}, \citenamefont {Urazhdin}, \citenamefont {Anane}, \citenamefont {Cros},\ and\ \citenamefont {Demokritov}}]{Demidov2020}%
  \BibitemOpen
  \bibfield  {author} {\bibinfo {author} {\bibfnamefont {V.}~\bibnamefont {Demidov}}, \bibinfo {author} {\bibfnamefont {S.}~\bibnamefont {Urazhdin}}, \bibinfo {author} {\bibfnamefont {A.}~\bibnamefont {Anane}}, \bibinfo {author} {\bibfnamefont {V.}~\bibnamefont {Cros}},\ and\ \bibinfo {author} {\bibfnamefont {S.}~\bibnamefont {Demokritov}},\ }\bibfield  {title} {\bibinfo {title} {Spin--orbit-torque magnonics},\ }\href@noop {} {\bibfield  {journal} {\bibinfo  {journal} {Journal of Applied Physics}\ }\textbf {\bibinfo {volume} {127}} (\bibinfo {year} {2020})}\BibitemShut {NoStop}%
\bibitem [{\citenamefont {Wang}\ and\ \citenamefont {Roychowdhury}(2019)}]{wang2019oim}%
  \BibitemOpen
  \bibfield  {author} {\bibinfo {author} {\bibfnamefont {T.}~\bibnamefont {Wang}}\ and\ \bibinfo {author} {\bibfnamefont {J.}~\bibnamefont {Roychowdhury}},\ }\bibfield  {title} {\bibinfo {title} {{OIM: Oscillator-based Ising machines for solving combinatorial optimisation problems}},\ }in\ \href@noop {} {\emph {\bibinfo {booktitle} {International Conference on Unconventional Computation and Natural Computation}}}\ (\bibinfo {organization} {Springer},\ \bibinfo {year} {2019})\ pp.\ \bibinfo {pages} {232--256}\BibitemShut {NoStop}%
\bibitem [{\citenamefont {Makiuchi}\ \emph {et~al.}(2021)\citenamefont {Makiuchi}, \citenamefont {Hioki}, \citenamefont {Shimazu}, \citenamefont {Oikawa}, \citenamefont {Yokoi}, \citenamefont {Daimon},\ and\ \citenamefont {Saitoh}}]{makiuchi2021parametron}%
  \BibitemOpen
  \bibfield  {author} {\bibinfo {author} {\bibfnamefont {T.}~\bibnamefont {Makiuchi}}, \bibinfo {author} {\bibfnamefont {T.}~\bibnamefont {Hioki}}, \bibinfo {author} {\bibfnamefont {Y.}~\bibnamefont {Shimazu}}, \bibinfo {author} {\bibfnamefont {Y.}~\bibnamefont {Oikawa}}, \bibinfo {author} {\bibfnamefont {N.}~\bibnamefont {Yokoi}}, \bibinfo {author} {\bibfnamefont {S.}~\bibnamefont {Daimon}},\ and\ \bibinfo {author} {\bibfnamefont {E.}~\bibnamefont {Saitoh}},\ }\bibfield  {title} {\bibinfo {title} {{Parametron on magnetic dot: Stable and stochastic operation}},\ }\href@noop {} {\bibfield  {journal} {\bibinfo  {journal} {Applied Physics Letters}\ }\textbf {\bibinfo {volume} {118}},\ \bibinfo {pages} {022402} (\bibinfo {year} {2021})}\BibitemShut {NoStop}%
\bibitem [{\citenamefont {Elyasi}\ \emph {et~al.}(2022)\citenamefont {Elyasi}, \citenamefont {Saitoh},\ and\ \citenamefont {Bauer}}]{elyasi2022stochasticity}%
  \BibitemOpen
  \bibfield  {author} {\bibinfo {author} {\bibfnamefont {M.}~\bibnamefont {Elyasi}}, \bibinfo {author} {\bibfnamefont {E.}~\bibnamefont {Saitoh}},\ and\ \bibinfo {author} {\bibfnamefont {G.~E.}\ \bibnamefont {Bauer}},\ }\bibfield  {title} {\bibinfo {title} {{Stochasticity of the magnon parametron}},\ }\href@noop {} {\bibfield  {journal} {\bibinfo  {journal} {Physical Review B}\ }\textbf {\bibinfo {volume} {105}},\ \bibinfo {pages} {054403} (\bibinfo {year} {2022})}\BibitemShut {NoStop}%
\bibitem [{\citenamefont {Sharma}\ \emph {et~al.}(2021)\citenamefont {Sharma}, \citenamefont {Mishra}, \citenamefont {Ngo}, \citenamefont {Guo}, \citenamefont {Fukami}, \citenamefont {Sato}, \citenamefont {Ohno},\ and\ \citenamefont {Yang}}]{Sharma2021}%
  \BibitemOpen
  \bibfield  {author} {\bibinfo {author} {\bibfnamefont {R.}~\bibnamefont {Sharma}}, \bibinfo {author} {\bibfnamefont {R.}~\bibnamefont {Mishra}}, \bibinfo {author} {\bibfnamefont {T.}~\bibnamefont {Ngo}}, \bibinfo {author} {\bibfnamefont {Y.-X.}\ \bibnamefont {Guo}}, \bibinfo {author} {\bibfnamefont {S.}~\bibnamefont {Fukami}}, \bibinfo {author} {\bibfnamefont {H.}~\bibnamefont {Sato}}, \bibinfo {author} {\bibfnamefont {H.}~\bibnamefont {Ohno}},\ and\ \bibinfo {author} {\bibfnamefont {H.}~\bibnamefont {Yang}},\ }\bibfield  {title} {\bibinfo {title} {Electrically connected spin-torque oscillators array for 2.4 ghz wifi band transmission and energy harvesting},\ }\href@noop {} {\bibfield  {journal} {\bibinfo  {journal} {Nature Communications}\ }\textbf {\bibinfo {volume} {12}},\ \bibinfo {pages} {2924} (\bibinfo {year} {2021})}\BibitemShut {NoStop}%
\bibitem [{\citenamefont {Litvinenko}\ \emph {et~al.}(2022)\citenamefont {Litvinenko}, \citenamefont {Sidi El~Valli}, \citenamefont {Iurchuk}, \citenamefont {Louis}, \citenamefont {Tyberkevych}, \citenamefont {Dieny}, \citenamefont {Slavin},\ and\ \citenamefont {Ebels}}]{Litvinenko2022}%
  \BibitemOpen
  \bibfield  {author} {\bibinfo {author} {\bibfnamefont {A.}~\bibnamefont {Litvinenko}}, \bibinfo {author} {\bibfnamefont {A.}~\bibnamefont {Sidi El~Valli}}, \bibinfo {author} {\bibfnamefont {V.}~\bibnamefont {Iurchuk}}, \bibinfo {author} {\bibfnamefont {S.}~\bibnamefont {Louis}}, \bibinfo {author} {\bibfnamefont {V.}~\bibnamefont {Tyberkevych}}, \bibinfo {author} {\bibfnamefont {B.}~\bibnamefont {Dieny}}, \bibinfo {author} {\bibfnamefont {A.~N.}\ \bibnamefont {Slavin}},\ and\ \bibinfo {author} {\bibfnamefont {U.}~\bibnamefont {Ebels}},\ }\bibfield  {title} {\bibinfo {title} {Ultrafast ghz-range swept-tuned spectrum analyzer with 20 ns temporal resolution based on a spin-torque nano-oscillator with a uniformly magnetized “free” layer},\ }\href@noop {} {\bibfield  {journal} {\bibinfo  {journal} {Nano Letters}\ }\textbf {\bibinfo {volume} {22}},\ \bibinfo {pages} {1874} (\bibinfo {year} {2022})}\BibitemShut {NoStop}%
\bibitem [{\citenamefont {Locatelli}\ \emph {et~al.}(2014)\citenamefont {Locatelli}, \citenamefont {Cros},\ and\ \citenamefont {Grollier}}]{Locatelli2014}%
  \BibitemOpen
  \bibfield  {author} {\bibinfo {author} {\bibfnamefont {N.}~\bibnamefont {Locatelli}}, \bibinfo {author} {\bibfnamefont {V.}~\bibnamefont {Cros}},\ and\ \bibinfo {author} {\bibfnamefont {J.}~\bibnamefont {Grollier}},\ }\bibfield  {title} {\bibinfo {title} {{Spin-torque building blocks}},\ }\href@noop {} {\bibfield  {journal} {\bibinfo  {journal} {Nature Mater}\ }\textbf {\bibinfo {volume} {13}},\ \bibinfo {pages} {11} (\bibinfo {year} {2014})}\BibitemShut {NoStop}%
\bibitem [{\citenamefont {Torrejon}\ \emph {et~al.}(2017)\citenamefont {Torrejon}, \citenamefont {Riou}, \citenamefont {Araujo}, \citenamefont {Tsunegi}, \citenamefont {Khalsa}, \citenamefont {Querlioz}, \citenamefont {Bortolotti}, \citenamefont {Cros}, \citenamefont {Yakushiji}, \citenamefont {Fukushima}, \citenamefont {Kubota}, \citenamefont {Yuasa}, \citenamefont {Stiles},\ and\ \citenamefont {Grollier}}]{Torrejon2017}%
  \BibitemOpen
  \bibfield  {author} {\bibinfo {author} {\bibfnamefont {J.}~\bibnamefont {Torrejon}}, \bibinfo {author} {\bibfnamefont {M.}~\bibnamefont {Riou}}, \bibinfo {author} {\bibfnamefont {F.~A.}\ \bibnamefont {Araujo}}, \bibinfo {author} {\bibfnamefont {S.}~\bibnamefont {Tsunegi}}, \bibinfo {author} {\bibfnamefont {G.}~\bibnamefont {Khalsa}}, \bibinfo {author} {\bibfnamefont {D.}~\bibnamefont {Querlioz}}, \bibinfo {author} {\bibfnamefont {P.}~\bibnamefont {Bortolotti}}, \bibinfo {author} {\bibfnamefont {V.}~\bibnamefont {Cros}}, \bibinfo {author} {\bibfnamefont {K.}~\bibnamefont {Yakushiji}}, \bibinfo {author} {\bibfnamefont {A.}~\bibnamefont {Fukushima}}, \bibinfo {author} {\bibfnamefont {H.}~\bibnamefont {Kubota}}, \bibinfo {author} {\bibfnamefont {S.}~\bibnamefont {Yuasa}}, \bibinfo {author} {\bibfnamefont {M.~D.}\ \bibnamefont {Stiles}},\ and\ \bibinfo {author} {\bibfnamefont {J.}~\bibnamefont {Grollier}},\ }\bibfield  {title} {\bibinfo {title} {{Neuromorphic computing with nanoscale spintronic oscillators}},\
  }\href@noop {} {\bibfield  {journal} {\bibinfo  {journal} {Nature}\ }\textbf {\bibinfo {volume} {547}},\ \bibinfo {pages} {428} (\bibinfo {year} {2017})}\BibitemShut {NoStop}%
\bibitem [{\citenamefont {Tsunegi}\ \emph {et~al.}(2019)\citenamefont {Tsunegi}, \citenamefont {Taniguchi}, \citenamefont {Nakajima}, \citenamefont {Miwa}, \citenamefont {Yakushiji}, \citenamefont {Fukushima}, \citenamefont {Yuasa},\ and\ \citenamefont {Kubota}}]{Tsunegi2019}%
  \BibitemOpen
  \bibfield  {author} {\bibinfo {author} {\bibfnamefont {S.}~\bibnamefont {Tsunegi}}, \bibinfo {author} {\bibfnamefont {T.}~\bibnamefont {Taniguchi}}, \bibinfo {author} {\bibfnamefont {K.}~\bibnamefont {Nakajima}}, \bibinfo {author} {\bibfnamefont {S.}~\bibnamefont {Miwa}}, \bibinfo {author} {\bibfnamefont {K.}~\bibnamefont {Yakushiji}}, \bibinfo {author} {\bibfnamefont {A.}~\bibnamefont {Fukushima}}, \bibinfo {author} {\bibfnamefont {S.}~\bibnamefont {Yuasa}},\ and\ \bibinfo {author} {\bibfnamefont {H.}~\bibnamefont {Kubota}},\ }\bibfield  {title} {\bibinfo {title} {Physical reservoir computing based on spin torque oscillator with forced synchronization},\ }\href@noop {} {\bibfield  {journal} {\bibinfo  {journal} {Applied Physics Letters}\ }\textbf {\bibinfo {volume} {114}} (\bibinfo {year} {2019})}\BibitemShut {NoStop}%
\bibitem [{\citenamefont {Zahedinejad}\ \emph {et~al.}(2019)\citenamefont {Zahedinejad}, \citenamefont {Awad}, \citenamefont {Muralidhar}, \citenamefont {Khymyn}, \citenamefont {Fulara}, \citenamefont {Mazraati}, \citenamefont {Dvornik},\ and\ \citenamefont {Akerman}}]{Zahedinejad2019}%
  \BibitemOpen
  \bibfield  {author} {\bibinfo {author} {\bibfnamefont {M.}~\bibnamefont {Zahedinejad}}, \bibinfo {author} {\bibfnamefont {A.~A.}\ \bibnamefont {Awad}}, \bibinfo {author} {\bibfnamefont {S.}~\bibnamefont {Muralidhar}}, \bibinfo {author} {\bibfnamefont {R.}~\bibnamefont {Khymyn}}, \bibinfo {author} {\bibfnamefont {H.}~\bibnamefont {Fulara}}, \bibinfo {author} {\bibfnamefont {H.}~\bibnamefont {Mazraati}}, \bibinfo {author} {\bibfnamefont {M.}~\bibnamefont {Dvornik}},\ and\ \bibinfo {author} {\bibfnamefont {J.}~\bibnamefont {Akerman}},\ }\bibfield  {title} {\bibinfo {title} {{Two-dimensional mutually synchronized spin Hall nano-oscillator arrays for neuromorphic computing}},\ }\href@noop {} {\bibfield  {journal} {\bibinfo  {journal} {Nat. Nanotechnol.}\ }\textbf {\bibinfo {volume} {15}},\ \bibinfo {pages} {47} (\bibinfo {year} {2019})}\BibitemShut {NoStop}%
\bibitem [{\citenamefont {Romera}\ \emph {et~al.}(2022)\citenamefont {Romera}, \citenamefont {Talatchian}, \citenamefont {Tsunegi}, \citenamefont {Yakushiji}, \citenamefont {Fukushima}, \citenamefont {Kubota}, \citenamefont {Yuasa}, \citenamefont {Cros}, \citenamefont {Bortolotti}, \citenamefont {Ernoult} \emph {et~al.}}]{Romera2022}%
  \BibitemOpen
  \bibfield  {author} {\bibinfo {author} {\bibfnamefont {M.}~\bibnamefont {Romera}}, \bibinfo {author} {\bibfnamefont {P.}~\bibnamefont {Talatchian}}, \bibinfo {author} {\bibfnamefont {S.}~\bibnamefont {Tsunegi}}, \bibinfo {author} {\bibfnamefont {K.}~\bibnamefont {Yakushiji}}, \bibinfo {author} {\bibfnamefont {A.}~\bibnamefont {Fukushima}}, \bibinfo {author} {\bibfnamefont {H.}~\bibnamefont {Kubota}}, \bibinfo {author} {\bibfnamefont {S.}~\bibnamefont {Yuasa}}, \bibinfo {author} {\bibfnamefont {V.}~\bibnamefont {Cros}}, \bibinfo {author} {\bibfnamefont {P.}~\bibnamefont {Bortolotti}}, \bibinfo {author} {\bibfnamefont {M.}~\bibnamefont {Ernoult}}, \emph {et~al.},\ }\bibfield  {title} {\bibinfo {title} {Binding events through the mutual synchronization of spintronic nano-neurons},\ }\href@noop {} {\bibfield  {journal} {\bibinfo  {journal} {Nature Communications}\ }\textbf {\bibinfo {volume} {13}},\ \bibinfo {pages} {883} (\bibinfo {year} {2022})}\BibitemShut {NoStop}%
\bibitem [{\citenamefont {Markovi\ifmmode~\acute{c}\else \'{c}\fi{}}\ \emph {et~al.}(2022)\citenamefont {Markovi\ifmmode~\acute{c}\else \'{c}\fi{}}, \citenamefont {Daniels}, \citenamefont {Sethi}, \citenamefont {Kent}, \citenamefont {Stiles},\ and\ \citenamefont {Grollier}}]{Marković2022}%
  \BibitemOpen
  \bibfield  {author} {\bibinfo {author} {\bibfnamefont {D.}~\bibnamefont {Markovi\ifmmode~\acute{c}\else \'{c}\fi{}}}, \bibinfo {author} {\bibfnamefont {M.~W.}\ \bibnamefont {Daniels}}, \bibinfo {author} {\bibfnamefont {P.}~\bibnamefont {Sethi}}, \bibinfo {author} {\bibfnamefont {A.~D.}\ \bibnamefont {Kent}}, \bibinfo {author} {\bibfnamefont {M.~D.}\ \bibnamefont {Stiles}},\ and\ \bibinfo {author} {\bibfnamefont {J.}~\bibnamefont {Grollier}},\ }\bibfield  {title} {\bibinfo {title} {Easy-plane spin hall nano-oscillators as spiking neurons for neuromorphic computing},\ }\href {https://doi.org/10.1103/PhysRevB.105.014411} {\bibfield  {journal} {\bibinfo  {journal} {Phys. Rev. B}\ }\textbf {\bibinfo {volume} {105}},\ \bibinfo {pages} {014411} (\bibinfo {year} {2022})}\BibitemShut {NoStop}%
\bibitem [{\citenamefont {Kurebayashi}\ \emph {et~al.}(2011)\citenamefont {Kurebayashi}, \citenamefont {Dzyapko}, \citenamefont {Demidov}, \citenamefont {Fang}, \citenamefont {Ferguson},\ and\ \citenamefont {Demokritov}}]{Kurebayashi2010}%
  \BibitemOpen
  \bibfield  {author} {\bibinfo {author} {\bibfnamefont {H.}~\bibnamefont {Kurebayashi}}, \bibinfo {author} {\bibfnamefont {O.}~\bibnamefont {Dzyapko}}, \bibinfo {author} {\bibfnamefont {V.}~\bibnamefont {Demidov}}, \bibinfo {author} {\bibfnamefont {D.}~\bibnamefont {Fang}}, \bibinfo {author} {\bibfnamefont {A.~J.}\ \bibnamefont {Ferguson}},\ and\ \bibinfo {author} {\bibfnamefont {S.}~\bibnamefont {Demokritov}},\ }\bibfield  {title} {\bibinfo {title} {{Controlled enhancement of spin-current emission by three-magnon splitting}},\ }\href@noop {} {\bibfield  {journal} {\bibinfo  {journal} {Nature Mater}\ }\textbf {\bibinfo {volume} {10}},\ \bibinfo {pages} {660–664} (\bibinfo {year} {2011})}\BibitemShut {NoStop}%
\bibitem [{\citenamefont {Balinsky}\ \emph {et~al.}(2016)\citenamefont {Balinsky}, \citenamefont {Haidar}, \citenamefont {Ranjbar}, \citenamefont {D{\"u}rrenfeld}, \citenamefont {Houshang}, \citenamefont {Slavin},\ and\ \citenamefont {{\AA}kerman}}]{balinsky2016}%
  \BibitemOpen
  \bibfield  {author} {\bibinfo {author} {\bibfnamefont {M.}~\bibnamefont {Balinsky}}, \bibinfo {author} {\bibfnamefont {M.}~\bibnamefont {Haidar}}, \bibinfo {author} {\bibfnamefont {M.}~\bibnamefont {Ranjbar}}, \bibinfo {author} {\bibfnamefont {P.}~\bibnamefont {D{\"u}rrenfeld}}, \bibinfo {author} {\bibfnamefont {A.}~\bibnamefont {Houshang}}, \bibinfo {author} {\bibfnamefont {A.}~\bibnamefont {Slavin}},\ and\ \bibinfo {author} {\bibfnamefont {J.}~\bibnamefont {{\AA}kerman}},\ }\bibfield  {title} {\bibinfo {title} {Modulation of the spectral characteristics of a nano-contact spin-torque oscillator via spin waves in an adjacent yttrium-iron garnet film},\ }\href@noop {} {\bibfield  {journal} {\bibinfo  {journal} {IEEE Magnetics Letters}\ }\textbf {\bibinfo {volume} {7}},\ \bibinfo {pages} {1} (\bibinfo {year} {2016})}\BibitemShut {NoStop}%
\bibitem [{\citenamefont {Safranski}\ \emph {et~al.}(2017)\citenamefont {Safranski}, \citenamefont {Barsukov}, \citenamefont {Lee}, \citenamefont {Schneider}, \citenamefont {Jara}, \citenamefont {Smith}, \citenamefont {Chang}, \citenamefont {Lenz}, \citenamefont {Lindner}, \citenamefont {Tserkovnyak} \emph {et~al.}}]{safranski2017}%
  \BibitemOpen
  \bibfield  {author} {\bibinfo {author} {\bibfnamefont {C.}~\bibnamefont {Safranski}}, \bibinfo {author} {\bibfnamefont {I.}~\bibnamefont {Barsukov}}, \bibinfo {author} {\bibfnamefont {H.~K.}\ \bibnamefont {Lee}}, \bibinfo {author} {\bibfnamefont {T.}~\bibnamefont {Schneider}}, \bibinfo {author} {\bibfnamefont {A.}~\bibnamefont {Jara}}, \bibinfo {author} {\bibfnamefont {A.}~\bibnamefont {Smith}}, \bibinfo {author} {\bibfnamefont {H.}~\bibnamefont {Chang}}, \bibinfo {author} {\bibfnamefont {K.}~\bibnamefont {Lenz}}, \bibinfo {author} {\bibfnamefont {J.}~\bibnamefont {Lindner}}, \bibinfo {author} {\bibfnamefont {Y.}~\bibnamefont {Tserkovnyak}}, \emph {et~al.},\ }\bibfield  {title} {\bibinfo {title} {Spin caloritronic nano-oscillator},\ }\href@noop {} {\bibfield  {journal} {\bibinfo  {journal} {Nature Communications}\ }\textbf {\bibinfo {volume} {8}},\ \bibinfo {pages} {117} (\bibinfo {year} {2017})}\BibitemShut {NoStop}%
\bibitem [{\citenamefont {Evelt}\ \emph {et~al.}(2018{\natexlab{a}})\citenamefont {Evelt}, \citenamefont {Safranski}, \citenamefont {Aldosary}, \citenamefont {Demidov}, \citenamefont {Barsukov}, \citenamefont {Nosov}, \citenamefont {Rinkevich}, \citenamefont {Sobotkiewich}, \citenamefont {Li}, \citenamefont {Shi} \emph {et~al.}}]{Evelt2018}%
  \BibitemOpen
  \bibfield  {author} {\bibinfo {author} {\bibfnamefont {M.}~\bibnamefont {Evelt}}, \bibinfo {author} {\bibfnamefont {C.}~\bibnamefont {Safranski}}, \bibinfo {author} {\bibfnamefont {M.}~\bibnamefont {Aldosary}}, \bibinfo {author} {\bibfnamefont {V.}~\bibnamefont {Demidov}}, \bibinfo {author} {\bibfnamefont {I.}~\bibnamefont {Barsukov}}, \bibinfo {author} {\bibfnamefont {A.}~\bibnamefont {Nosov}}, \bibinfo {author} {\bibfnamefont {A.}~\bibnamefont {Rinkevich}}, \bibinfo {author} {\bibfnamefont {K.}~\bibnamefont {Sobotkiewich}}, \bibinfo {author} {\bibfnamefont {X.}~\bibnamefont {Li}}, \bibinfo {author} {\bibfnamefont {J.}~\bibnamefont {Shi}}, \emph {et~al.},\ }\bibfield  {title} {\bibinfo {title} {Spin hall-induced auto-oscillations in ultrathin yig grown on pt},\ }\href@noop {} {\bibfield  {journal} {\bibinfo  {journal} {Scientific Reports}\ }\textbf {\bibinfo {volume} {8}},\ \bibinfo {pages} {1269} (\bibinfo {year} {2018}{\natexlab{a}})}\BibitemShut {NoStop}%
\bibitem [{\citenamefont {Kajiwara}\ \emph {et~al.}(2010)\citenamefont {Kajiwara}, \citenamefont {Harii}, \citenamefont {Takahashi}, \citenamefont {Ohe}, \citenamefont {Uchida}, \citenamefont {Mizuguchi}, \citenamefont {Umezawa}, \citenamefont {Kawai}, \citenamefont {Ando}, \citenamefont {Takanashi}, \citenamefont {Maekawa},\ and\ \citenamefont {Saitoh}}]{Kajiwara2010}%
  \BibitemOpen
  \bibfield  {author} {\bibinfo {author} {\bibfnamefont {Y.}~\bibnamefont {Kajiwara}}, \bibinfo {author} {\bibfnamefont {K.}~\bibnamefont {Harii}}, \bibinfo {author} {\bibfnamefont {S.}~\bibnamefont {Takahashi}}, \bibinfo {author} {\bibfnamefont {J.}~\bibnamefont {Ohe}}, \bibinfo {author} {\bibfnamefont {K.}~\bibnamefont {Uchida}}, \bibinfo {author} {\bibfnamefont {M.}~\bibnamefont {Mizuguchi}}, \bibinfo {author} {\bibfnamefont {H.}~\bibnamefont {Umezawa}}, \bibinfo {author} {\bibfnamefont {H.}~\bibnamefont {Kawai}}, \bibinfo {author} {\bibfnamefont {K.}~\bibnamefont {Ando}}, \bibinfo {author} {\bibfnamefont {K.}~\bibnamefont {Takanashi}}, \bibinfo {author} {\bibfnamefont {S.}~\bibnamefont {Maekawa}},\ and\ \bibinfo {author} {\bibfnamefont {E.}~\bibnamefont {Saitoh}},\ }\bibfield  {title} {\bibinfo {title} {{Transmission of electrical signals by spin-wave interconversion in a magnetic insulator}},\ }\href@noop {} {\bibfield  {journal} {\bibinfo  {journal} {Nature}\ }\textbf {\bibinfo {volume} {464}},\
  \bibinfo {pages} {262} (\bibinfo {year} {2010})}\BibitemShut {NoStop}%
\bibitem [{\citenamefont {Zheng}\ \emph {et~al.}(2024)\citenamefont {Zheng}, \citenamefont {Channa}, \citenamefont {Riddifor}, \citenamefont {Wisser}, \citenamefont {Mahalingam}, \citenamefont {Bowers}, \citenamefont {McConney}, \citenamefont {N'Diaye}, \citenamefont {Vailionis}, \citenamefont {Cogulu}, \citenamefont {Ren}, \citenamefont {Galazka}, \citenamefont {Kent},\ and\ \citenamefont {Suzuki}}]{Zheng2024}%
  \BibitemOpen
  \bibfield  {author} {\bibinfo {author} {\bibfnamefont {X.~Y.}\ \bibnamefont {Zheng}}, \bibinfo {author} {\bibfnamefont {S.}~\bibnamefont {Channa}}, \bibinfo {author} {\bibfnamefont {L.~J.}\ \bibnamefont {Riddifor}}, \bibinfo {author} {\bibfnamefont {J.~J.}\ \bibnamefont {Wisser}}, \bibinfo {author} {\bibfnamefont {K.}~\bibnamefont {Mahalingam}}, \bibinfo {author} {\bibfnamefont {C.~T.}\ \bibnamefont {Bowers}}, \bibinfo {author} {\bibfnamefont {M.~E.}\ \bibnamefont {McConney}}, \bibinfo {author} {\bibfnamefont {A.~T.}\ \bibnamefont {N'Diaye}}, \bibinfo {author} {\bibfnamefont {A.}~\bibnamefont {Vailionis}}, \bibinfo {author} {\bibfnamefont {E.}~\bibnamefont {Cogulu}}, \bibinfo {author} {\bibfnamefont {H.}~\bibnamefont {Ren}}, \bibinfo {author} {\bibfnamefont {Z.}~\bibnamefont {Galazka}}, \bibinfo {author} {\bibfnamefont {A.~D.}\ \bibnamefont {Kent}},\ and\ \bibinfo {author} {\bibfnamefont {Y.}~\bibnamefont {Suzuki}},\ }\bibfield  {title} {\bibinfo {title} {{Ultra-thin lithium aluminate spinel ferrite films
  with perpendicular magnetic anisotropy and low damping}},\ }\href@noop {} {\bibfield  {journal} {\bibinfo  {journal} {Nat. Commun.}\ }\textbf {\bibinfo {volume} {14}},\ \bibinfo {pages} {4918} (\bibinfo {year} {2024})}\BibitemShut {NoStop}%
\bibitem [{\citenamefont {Ren}\ \emph {et~al.}(2023)\citenamefont {Ren}, \citenamefont {Zheng}, \citenamefont {Channa}, \citenamefont {Wu}, \citenamefont {O’Mahoney}, \citenamefont {Suzuki},\ and\ \citenamefont {Kent}}]{Ren2023}%
  \BibitemOpen
  \bibfield  {author} {\bibinfo {author} {\bibfnamefont {H.}~\bibnamefont {Ren}}, \bibinfo {author} {\bibfnamefont {X.~Y.}\ \bibnamefont {Zheng}}, \bibinfo {author} {\bibfnamefont {S.}~\bibnamefont {Channa}}, \bibinfo {author} {\bibfnamefont {G.}~\bibnamefont {Wu}}, \bibinfo {author} {\bibfnamefont {D.~A.}\ \bibnamefont {O’Mahoney}}, \bibinfo {author} {\bibfnamefont {Y.}~\bibnamefont {Suzuki}},\ and\ \bibinfo {author} {\bibfnamefont {A.~D.}\ \bibnamefont {Kent}},\ }\bibfield  {title} {\bibinfo {title} {{Hybrid spin Hall nano-oscillators based on ferromagnetic metal/ferrimagnetic insulator heterostructures}},\ }\href@noop {} {\bibfield  {journal} {\bibinfo  {journal} {Nat. Commun.}\ }\textbf {\bibinfo {volume} {14}},\ \bibinfo {pages} {1406} (\bibinfo {year} {2023})}\BibitemShut {NoStop}%
\bibitem [{\citenamefont {Evelt}\ \emph {et~al.}(2018{\natexlab{b}})\citenamefont {Evelt}, \citenamefont {Soumah}, \citenamefont {Rinkevich}, \citenamefont {Demokritov}, \citenamefont {Anane}, \citenamefont {Cros}, \citenamefont {Ben~Youssef}, \citenamefont {de~Loubens}, \citenamefont {Klein}, \citenamefont {Bortolotti},\ and\ \citenamefont {Demidov}}]{evelt2018emission}%
  \BibitemOpen
  \bibfield  {author} {\bibinfo {author} {\bibfnamefont {M.}~\bibnamefont {Evelt}}, \bibinfo {author} {\bibfnamefont {L.}~\bibnamefont {Soumah}}, \bibinfo {author} {\bibfnamefont {A.}~\bibnamefont {Rinkevich}}, \bibinfo {author} {\bibfnamefont {S.}~\bibnamefont {Demokritov}}, \bibinfo {author} {\bibfnamefont {A.}~\bibnamefont {Anane}}, \bibinfo {author} {\bibfnamefont {V.}~\bibnamefont {Cros}}, \bibinfo {author} {\bibfnamefont {J.}~\bibnamefont {Ben~Youssef}}, \bibinfo {author} {\bibfnamefont {G.}~\bibnamefont {de~Loubens}}, \bibinfo {author} {\bibfnamefont {O.}~\bibnamefont {Klein}}, \bibinfo {author} {\bibfnamefont {P.}~\bibnamefont {Bortolotti}},\ and\ \bibinfo {author} {\bibfnamefont {V.}~\bibnamefont {Demidov}},\ }\bibfield  {title} {\bibinfo {title} {Emission of coherent propagating magnons by insulator-based spin-orbit-torque oscillators},\ }\href@noop {} {\bibfield  {journal} {\bibinfo  {journal} {Physical Review Applied}\ }\textbf {\bibinfo {volume} {10}},\ \bibinfo {pages} {041002} (\bibinfo {year}
  {2018}{\natexlab{b}})}\BibitemShut {NoStop}%
\bibitem [{\citenamefont {Chen}\ \emph {et~al.}(2023)\citenamefont {Chen}, \citenamefont {Zhou}, \citenamefont {Tao}, \citenamefont {Gao}, \citenamefont {Liang}, \citenamefont {Li},\ and\ \citenamefont {Liu}}]{chen2023self}%
  \BibitemOpen
  \bibfield  {author} {\bibinfo {author} {\bibfnamefont {L.}~\bibnamefont {Chen}}, \bibinfo {author} {\bibfnamefont {K.}~\bibnamefont {Zhou}}, \bibinfo {author} {\bibfnamefont {Z.}~\bibnamefont {Tao}}, \bibinfo {author} {\bibfnamefont {Z.}~\bibnamefont {Gao}}, \bibinfo {author} {\bibfnamefont {L.}~\bibnamefont {Liang}}, \bibinfo {author} {\bibfnamefont {Z.}~\bibnamefont {Li}},\ and\ \bibinfo {author} {\bibfnamefont {R.}~\bibnamefont {Liu}},\ }\bibfield  {title} {\bibinfo {title} {Self-oscillations in a nanogap spin hall nano-oscillator with a perpendicularly magnetized external film},\ }\href@noop {} {\bibfield  {journal} {\bibinfo  {journal} {Physical Review Applied}\ }\textbf {\bibinfo {volume} {19}},\ \bibinfo {pages} {054051} (\bibinfo {year} {2023})}\BibitemShut {NoStop}%
\bibitem [{\citenamefont {Vansteenkiste}\ \emph {et~al.}(2014)\citenamefont {Vansteenkiste}, \citenamefont {Leliaert}, \citenamefont {Dvornik}, \citenamefont {Helsen}, \citenamefont {Garcia-Sanchez},\ and\ \citenamefont {Van~Waeyenberge}}]{Vansteenkiste2014}%
  \BibitemOpen
  \bibfield  {author} {\bibinfo {author} {\bibfnamefont {A.}~\bibnamefont {Vansteenkiste}}, \bibinfo {author} {\bibfnamefont {J.}~\bibnamefont {Leliaert}}, \bibinfo {author} {\bibfnamefont {M.}~\bibnamefont {Dvornik}}, \bibinfo {author} {\bibfnamefont {M.}~\bibnamefont {Helsen}}, \bibinfo {author} {\bibfnamefont {F.}~\bibnamefont {Garcia-Sanchez}},\ and\ \bibinfo {author} {\bibfnamefont {B.}~\bibnamefont {Van~Waeyenberge}},\ }\bibfield  {title} {\bibinfo {title} {{The design and verification of MuMax3}},\ }\href@noop {} {\bibfield  {journal} {\bibinfo  {journal} {AIP Advances}\ }\textbf {\bibinfo {volume} {4}},\ \bibinfo {pages} {107133} (\bibinfo {year} {2014})}\BibitemShut {NoStop}%
\bibitem [{\citenamefont {McMichael}\ and\ \citenamefont {Stiles}(2005)}]{McMichael2005}%
  \BibitemOpen
  \bibfield  {author} {\bibinfo {author} {\bibfnamefont {R.~D.}\ \bibnamefont {McMichael}}\ and\ \bibinfo {author} {\bibfnamefont {M.~D.}\ \bibnamefont {Stiles}},\ }\bibfield  {title} {\bibinfo {title} {{Magnetic normal modes of nanoelements}},\ }\href {https://doi.org/10.1063/1.1852191} {\bibfield  {journal} {\bibinfo  {journal} {Journal of Applied Physics}\ }\textbf {\bibinfo {volume} {97}},\ \bibinfo {pages} {10J901} (\bibinfo {year} {2005})}\BibitemShut {NoStop}%
\bibitem [{\citenamefont {Succar}\ and\ \citenamefont {Haidar}(2023)}]{succar2023}%
  \BibitemOpen
  \bibfield  {author} {\bibinfo {author} {\bibfnamefont {M.}~\bibnamefont {Succar}}\ and\ \bibinfo {author} {\bibfnamefont {M.}~\bibnamefont {Haidar}},\ }\bibfield  {title} {\bibinfo {title} {Spin wave excitations in a nanowire spin hall oscillator with perpendicular magnetic anisotropy},\ }\href@noop {} {\bibfield  {journal} {\bibinfo  {journal} {Journal of Applied Physics}\ }\textbf {\bibinfo {volume} {133}} (\bibinfo {year} {2023})}\BibitemShut {NoStop}%
\bibitem [{\citenamefont {Dvornik}\ \emph {et~al.}(2018)\citenamefont {Dvornik}, \citenamefont {Awad},\ and\ \citenamefont {{\AA}kerman}}]{dvornik2018origin}%
  \BibitemOpen
  \bibfield  {author} {\bibinfo {author} {\bibfnamefont {M.}~\bibnamefont {Dvornik}}, \bibinfo {author} {\bibfnamefont {A.~A.}\ \bibnamefont {Awad}},\ and\ \bibinfo {author} {\bibfnamefont {J.}~\bibnamefont {{\AA}kerman}},\ }\bibfield  {title} {\bibinfo {title} {Origin of magnetization auto-oscillations in constriction-based spin hall nano-oscillators},\ }\href@noop {} {\bibfield  {journal} {\bibinfo  {journal} {Physical Review Applied}\ }\textbf {\bibinfo {volume} {9}},\ \bibinfo {pages} {014017} (\bibinfo {year} {2018})}\BibitemShut {NoStop}%
\bibitem [{\citenamefont {Fan}\ \emph {et~al.}(2021)\citenamefont {Fan}, \citenamefont {Finley}, \citenamefont {Han}, \citenamefont {Holtz}, \citenamefont {Quarterman}, \citenamefont {Zhang}, \citenamefont {Safi}, \citenamefont {Hou}, \citenamefont {Grutter},\ and\ \citenamefont {Liu}}]{fan2021}%
  \BibitemOpen
  \bibfield  {author} {\bibinfo {author} {\bibfnamefont {Y.}~\bibnamefont {Fan}}, \bibinfo {author} {\bibfnamefont {J.}~\bibnamefont {Finley}}, \bibinfo {author} {\bibfnamefont {J.}~\bibnamefont {Han}}, \bibinfo {author} {\bibfnamefont {M.~E.}\ \bibnamefont {Holtz}}, \bibinfo {author} {\bibfnamefont {P.}~\bibnamefont {Quarterman}}, \bibinfo {author} {\bibfnamefont {P.}~\bibnamefont {Zhang}}, \bibinfo {author} {\bibfnamefont {T.~S.}\ \bibnamefont {Safi}}, \bibinfo {author} {\bibfnamefont {J.~T.}\ \bibnamefont {Hou}}, \bibinfo {author} {\bibfnamefont {A.~J.}\ \bibnamefont {Grutter}},\ and\ \bibinfo {author} {\bibfnamefont {L.}~\bibnamefont {Liu}},\ }\bibfield  {title} {\bibinfo {title} {Resonant spin transmission mediated by magnons in a magnetic insulator multilayer structure},\ }\href@noop {} {\bibfield  {journal} {\bibinfo  {journal} {Advanced Materials}\ }\textbf {\bibinfo {volume} {33}},\ \bibinfo {pages} {2008555} (\bibinfo {year} {2021})}\BibitemShut {NoStop}%
\end{thebibliography}%

\end{document}